\documentclass[pdflatex,sn-nature]{sn-jnl}
\usepackage{graphicx}%
\usepackage{multirow}%
\usepackage{amsmath,amssymb,amsfonts}%
\usepackage{amsthm}%
\usepackage{mathrsfs}%
\usepackage[title]{appendix}%
\usepackage{xcolor}%
\usepackage{textcomp}%
\usepackage{manyfoot}%
\usepackage{booktabs}%
\usepackage{algorithm}%
\usepackage{algorithmicx}%
\usepackage{algpseudocode}%
\usepackage{listings}%
\usepackage{comment}
\usepackage[version=4]{mhchem}
\usepackage{float}
\usepackage{tabularx}
\usepackage{rotating}
\usepackage{pifont}
\usepackage{setspace}
\begin{document}
\doublespacing

\title[Article Title]{The Precursor Genome: A Pairwise Reaction Dataset for Solid-State Synthesis}

\author[1,2]{\fnm{Lauren N.} \sur{Walters}}
\equalcont{These authors contributed equally to this work.}
\author[1]{\fnm{Matthew J.} \sur{McDermott}}
\equalcont{These authors contributed equally to this work.}
\author[3]{\fnm{Bernardus} \sur{Rendy}}
\author[3]{\fnm{Yuxing} \sur{Fei}}
\author[1,3]{\fnm{Kristin A.} \sur{Persson}}
\author*[1,3]{\fnm{Gerbrand} \sur{Ceder}}\email{gceder@berkeley.edu}

\affil[1]{\orgdiv{Division of Materials Science}, \orgname{Lawrence Berkeley National Laboratory}, \orgaddress{\street{1 Cyclotron Rd}, \city{Berkeley}, \postcode{94720}, \state{CA}, \country{U.S.A}}}

\affil*[2]{\orgdiv{Bakar Institute of Digital Materials for the Planet}, \orgname{University of California Berkeley}, \orgaddress{\street{373 Cory Hall}, \city{Berkeley}, \postcode{94720}, \state{CA}, \country{U.S.A}}}

\affil[3]{\orgdiv{Department of Materials Science and Engineering}, \orgname{University of California Berkeley}, \orgaddress{\street{210 Hearst Memorial Mining Building}, \city{Berkeley}, \postcode{94720}, \state{CA}, \country{U.S.A}}}

\abstract{

    Solid-state reactions remain the dominant route to inorganic materials, yet no large, machine-readable dataset reports their experimental protocols and outcomes with consistent provenance; this gap obstructs first-principles, data-driven, and machine-learning approaches to synthesis science. Here, we present the Precursor Genome, a dataset of 1,035 pairwise solid-state reactions generated autonomously by the A-Lab self-driving laboratory, spanning 46 precursors and 39 elements. Every reaction is reported together with its full experimental metadata, including measured thermal profiles, precursor and recovered masses, and instrument configuration. Every product mixture is identified from raw X-ray diffraction (1,351 scans) through automated Rietveld refinement with the Dara framework, yielding 1,950 refinement cases that are independently validated by human experts on a three-tier quality scale. Raw pattern files, serialized refinement objects, and reviewer annotations are distributed through a Pydantic-validated JSON ledger, preserving full traceability from each precursor pair to its final phase assignment. The Precursor Genome establishes a FAIR, reusable benchmark for training and evaluating predictive models of solid-state reactivity.
}

\maketitle

\section{Background \& Summary}

Materials discovery is transitioning from case-by-case synthesis to autonomous, data-driven workflows \cite{Alab, MacLeod2020_SDL1, Cakan2024_SDL2, Burger2020_SDL3, Tom2024_SDL-Review}. While computational methods now predict millions of thermodynamically stable compounds \cite{Jain2013_MaterialsProject, Merchant2023_DeepMindGnome, Zeni2025_MatterGen, Cavignac2025_Alexandria}, their experimental realization remains a persistent bottleneck \cite{Mallouk_SolidState_Review, Jansen2018_moreModernReview, Kim2017_TextMining}. The published literature on experimental synthesis is a poor substitute for systematic data: it overwhelmingly reports only successful routes, omits negative trials, and rarely preserves the sample-handling logs and raw characterization output needed for true reproducibility \cite{Raccuglia2016_DarkReactions, Jia2019_AnthropogenicBiases, Baibakova2025_BFO-TextMining}. Hence, to enable data-driven synthesis science, the community requires large-scale datasets that capture the full spectrum of reaction outcomes within a standardized, well-documented experimental framework.

Self-driving laboratories (SDLs) offer a direct remedy for these systemic reporting gaps \cite{Alab, MacLeod2020_SDL1, Cakan2024_SDL2, Burger2020_SDL3, Tom2024_SDL-Review}. By standardizing execution while automatically capturing sensor traces, imaging, and timestamps under machine-readable protocols, SDLs enable the high-throughput generation of high-fidelity synthesis datasets with a consistency that is unattainable in traditional laboratory settings.

In this work, we introduce the Precursor Genome, a systematic repository of 1,035 reaction outcomes between pairs of 46 common inorganic precursors spanning 39 elements, executed on the A-Lab platform \cite{Alab, Alab_os}. Heating was explicitly performed at short time points (1-hr dwell) to capture early intermediate phase formation as well as ``negative'' data (i.e., precursor pairs that failed or are slow to react). For every reaction coordinate, the dataset records raw XRD patterns, expert-validated Rietveld refinements, and high-fidelity experimental metadata, and is distributed under FAIR principles through a persistent identifier, an open license, a Pydantic-validated schema, and a controlled reaction-outcome vocabulary. By providing a transparent, machine-readable map of synthetic accessibility, the Precursor Genome establishes a benchmark for training and evaluating predictive algorithms in data-driven solid-state synthesis.

\section{Methods}

\subsection{Precursor Preparation and Testing}

We curated a precursor library designed to span a diverse chemical space (39 elements), prioritizing materials common to solid-state synthesis with an emphasis on elements for electrochemical energy storage. Selection criteria included precursor cost, stability in air, hygroscopicity, and handling safety. Reagents were purchased from Sigma-Aldrich (Millipore Sigma) at generally 99\% or greater purity. Table~\ref{tab:precursor-purchase-info} contains all precursors used, CAS numbers, manufacturer, and purity.

Six precursors with significant hygroscopicity and/or handling safety concerns (\ce{LiC2H3O2}$\cdot$\ce{2H2O}, \ce{(NH4)2SO4}, \ce{NH4Cl}, \ce{NaH2PO4}$\cdot$\ce{2H2O}, \ce{NaNO3}, and \ce{Si}) were eliminated during the data collection. The majority of precursors were used directly as is. However, some compounds required additional preprocessing (milling) to be compatible with the automated dosing heads in the A-Lab.

To maintain high data integrity, every precursor underwent rigorous pre-synthesis characterization by powder XRD, SEM, and EDS, together with an air-sensitivity screening. For the air-sensitivity screening, a nominal 0.5\,g sample of each precursor was weighed into an uncapped container and left open to ambient laboratory air at Lawrence Berkeley National Laboratory for 48\,h; the container and sample were then re-weighed, and the fractional mass change (``Air sens.'' in Table~\ref{tab:precursor-purchase-info}) was computed as $\Delta m / m_0 = [(\text{post-exposure mass}) - m_0] / m_0$. Detailed procurement, purity, air-sensitivity, and pre-processing information are compiled in Table~\ref{tab:precursor-purchase-info}.

\begin{sidewaystable}
    \centering
    \footnotesize
    \caption{\textbf{Precursor inventory and pre-processing.} All reagents were purchased from Sigma-Aldrich (SA), i.e.,MilliporeSigma. A dash (---) in the ``pre-processing'' column indicates the precursor was used as supplied; a dash (--) in ``used'' marks a precursor removed due to safety or hygroscopicity (greyed; not part of the pairwise campaign). Raw masses behind the ``Air sens.'' column are in the Supplementary Information. Purities are listed as provided by the supplier. Melting-point references: \textsuperscript{a}\,Lange's Handbook (15th Ed.); \textsuperscript{b}\,American Elements; \textsuperscript{c}\,Fisher Scientific; \textsuperscript{d}\,NIOSH, 2023; \textsuperscript{e}\,Sigma-Aldrich; \textsuperscript{f}\,Brown \& Gallagher (2008); \textsuperscript{g}\,CAS Common Chemistry; \textsuperscript{h}\,Haynes, CRC Handbook (95th Ed.); \textsuperscript{i}\,Weast, CRC Handbook (68th Ed.).}
    \label{tab:precursor-purchase-info}
    \begin{tabularx}{\linewidth}{@{}cc >{\raggedright\arraybackslash}p{5.2cm} lllrrcrX@{}}
        \toprule
        Index & Used      & Precursor Name                                            & Formula                                            & CAS Number                      & SA Prod.\ No.                        & Purity                     & $T_m$ ($^\circ$C)         & Ref.                                     & Air sens. (\%)             & Pre-processing                                \\
        \midrule
        1     & \ding{51} & silver(i) oxide                                           & \ce{Ag2O}                                          & 20667-12-3                      & \texttt{221163}                      & 0.99                       & 280                       & \textsuperscript{b}                      & +3.1                       & ---                                           \\
        2     & \ding{51} & aluminium hydroxide                                       & \ce{Al(OH)3}                                       & 21645-51-2                      & \texttt{1.01093}                     &                            & 300                       & \textsuperscript{a}                      & -0.06                      & ---                                           \\
        3     & \ding{51} & boric acid                                                & \ce{B(OH)3}                                        & 10043-35-3                      & \texttt{B0252}                       & 0.995                      & 171                       & \textsuperscript{a}                      & -0.23                      & Speedmix (dry): 5\,min, 3000\,rpm                    \\
        4     & \ding{51} & barium carbonate                                          & \ce{BaCO3}                                         & 513-77-9                        & \texttt{237108}                      & 0.99                       & 811                       & \textsuperscript{c}                      & +0.02                      & ---                                           \\
        5     & \ding{51} & barium peroxide                                           & \ce{BaO2}                                          & 1304-29-6                       & \texttt{769304}                      & 0.86                       & 450                       & \textsuperscript{a}                      & +0.46                      & ---                                           \\
        6     & \ding{51} & bismuth(iii) oxide                                        & \ce{Bi2O3}                                         & 1304-76-3                       & \texttt{223891}                      & 0.999                      & 817                       & \textsuperscript{a}                      & +0.06                      & ---                                           \\
        7     & \ding{51} & calcium carbonate                                         & \ce{CaCO3}                                         & 471-34-1                        & \texttt{239216}                      & 0.99                       & 825                       & \textsuperscript{a}                      & +0.10                      & ---                                           \\
        8     & \ding{51} & cobalt(ii) oxide                                          & \ce{CoO}                                           & 1307-96-6                       & \texttt{343153}                      &                            & 1935                      & \textsuperscript{a}                      & +0.08                      & ---                                           \\
        9     & \ding{51} & cobalt(ii,iii) oxide                                      & \ce{Co3O4}                                         & 1308-06-1                       & \texttt{221643}                      &                            & 895                       & \textsuperscript{e}                      & -0.65                      & ---                                           \\
        10    & \ding{51} & chromium(iii) oxide                                       & \ce{Cr2O3}                                         & 1308-38-9                       & \texttt{393703}                      & 0.98                       & 2435                      & \textsuperscript{b}                      & -0.05                      & ---                                           \\
        11    & \ding{51} & copper(ii) oxide                                          & \ce{CuO}                                           & 1317-38-0                       & \texttt{241741}                      & 0.99                       & 1326                      & \textsuperscript{i}                      & -0.04                      & ---                                           \\
        12    & \ding{51} & iron(ii,iii) oxide                                        & \ce{Fe3O4}                                         & 1317-61-9                       & \texttt{310069}                      & 0.95                       & 1597                      & \textsuperscript{a}                      & -0.26                      & ---                                           \\
        13    & \ding{51} & iron(iii) oxide                                           & \ce{Fe2O3}                                         & 1309-37-1                       & \texttt{310050}                      & 0.96                       & 1565                      & \textsuperscript{a}                      & +0.46                      & ---                                           \\
        14    & \ding{51} & iron(ii) oxalate dihydrate                                & \ce{FeC2O4}$\cdot$\ce{2H2O}                        & 6047-25-2                       & \texttt{307726}                      & 0.99                       & 190                       & \textsuperscript{b}                      & -0.29                      & ---                                           \\
        15    & \ding{51} & gallium(iii) oxide                                        & \ce{Ga2O3}                                         & 12024-21-4                      & \texttt{215066}                      & 0.9999                     & 1900                      & \textsuperscript{b}                      & -0.07                      & ---                                           \\
        16    & \ding{51} & germanium(iv) oxide                                       & \ce{GeO2}                                          & 1310-53-8                       & \texttt{199478}                      & 0.99998                    & 1115                      & \textsuperscript{a}                      & -0.05                      & ---                                           \\
        17    & \ding{51} & hafnium(iv) oxide                                         & \ce{HfO2}                                          & 12055-23-1                      & \texttt{202118}                      & 0.98                       & 2774                      & \textsuperscript{a}                      & -0.07                      & ---                                           \\
        18    & \ding{51} & indium(iii) oxide                                         & \ce{In2O3}                                         & 1312-43-2                       & \texttt{289418}                      & 0.9999                     & 1910                      & \textsuperscript{g}                      & -0.06                      & ---                                           \\
        19    & \ding{51} & potassium carbonate sesquihydrate                         & \ce{K2CO3}$\cdot$\ce{1.5H2O}                       & 6381-79-9                       & \texttt{243558}                      & 0.99                       & 891                       & \textsuperscript{c}                      & +16.8                      & Mortar \& pestle: 5\,min                       \\
        20    & \ding{51} & potassium phosphate monobasic                             & \ce{KH2PO4}                                        & 7778-77-0                       & \texttt{P0662}                       & 0.99                       & 253                       & \textsuperscript{b}                      & +0.01                      & Mortar \& pestle: 5\,min                       \\
        21    & \ding{51} & lanthanum(iii) hydroxide                                  & \ce{La(OH)3}                                       & 14507-19-8                      & \texttt{447226}                      & 0.999                      & 330                       & \textsuperscript{f}                      & +0.29                      & ---                                           \\
        22    & \ding{51} & lithium carbonate                                         & \ce{Li2CO3}                                        & 554-13-2                        & \texttt{255823}                      & 0.99                       & 720                       & \textsuperscript{a}                      & -0.05                      & Mortar \& pestle: 5\,min                       \\
        23    & --        & \textcolor{gray!70}{lithium acetate dihydrate}            & \textcolor{gray!70}{\ce{LiC2H3O2}$\cdot$\ce{2H2O}} & \textcolor{gray!70}{6108-17-4}  & \textcolor{gray!70}{\texttt{L6883}}  &                            & \textcolor{gray!70}{53}   & \textcolor{gray!70}{\textsuperscript{e}} & \textcolor{gray!70}{+26.2} & \textcolor{gray!70}{---}                      \\
        24    & \ding{51} & lithium hydroxide monohydrate                             & \ce{LiOH}$\cdot$\ce{H2O}                           & 1310-66-3                       & \texttt{402974}                      & 0.98                       & 462                       & \textsuperscript{b}                      & +28.5                      & High energy ball mill: 5\,min                  \\
        25    & \ding{51} & magnesium oxide                                           & \ce{MgO}                                           & 1309-48-4                       & \texttt{243388}                      & 0.97                       & 2852                      & \textsuperscript{b}                      & +3.0                       & ---                                           \\
        26    & \ding{51} & manganese(ii) oxide                                       & \ce{MnO}                                           & 1344-43-0                       & \texttt{377201}                      & 0.99                       & 1840                      & \textsuperscript{a}                      & -0.03                      & High energy ball mill: 10\,min                 \\
        27    & \ding{51} & manganese(ii,iii) oxide                                   & \ce{Mn3O4}                                         & 1317-35-7                       & \texttt{377473}                      & 0.97                       & 1567                      & \textsuperscript{a}                      & +0.05                      & ---                                           \\
        28    & \ding{51} & manganese(iii) oxide                                      & \ce{Mn2O3}                                         & 1317-34-6                       & \texttt{377457}                      & 0.99                       & 940                       & \textsuperscript{b}                      & -0.09                      & ---                                           \\
        29    & \ding{51} & manganese(iv) oxide                                       & \ce{MnO2}                                          & 1313-13-9                       & \texttt{243442}                      & 0.99                       & 535                       & \textsuperscript{b}                      & -0.05                      & Speedmix (dry): 30\,min, 3000\,rpm                \\
        30    & \ding{51} & molybdenum(vi) oxide                                      & \ce{MoO3}                                          & 1313-27-5                       & \texttt{267856}                      & 0.995                      & 795                       & \textsuperscript{b}                      & -0.04                      & ---                                           \\
        31    & \ding{51} & ammonium phosphate dibasic                                & \ce{(NH4)2HPO4}                                    & 7783-28-0                       & \texttt{215996}                      & 0.98                       & 155                       & \textsuperscript{a}                      & -0.11                      & Mortar \& pestle 5\,min                       \\
        32    & --        & \textcolor{gray!70}{ammonium sulfate}                     & \textcolor{gray!70}{\ce{(NH4)2SO4}}                & \textcolor{gray!70}{7783-20-2}  & \textcolor{gray!70}{\texttt{A4915}}  & \textcolor{gray!70}{0.99}  & \textcolor{gray!70}{280}  & \textcolor{gray!70}{\textsuperscript{a}} & \textcolor{gray!70}{-0.03} & \textcolor{gray!70}{---}                      \\
        33    & --        & \textcolor{gray!70}{ammonium chloride}                    & \textcolor{gray!70}{\ce{NH4Cl}}                    & \textcolor{gray!70}{12125-02-9} & \textcolor{gray!70}{\texttt{A4514}}  & \textcolor{gray!70}{0.995} & \textcolor{gray!70}{350}  & \textcolor{gray!70}{\textsuperscript{d}} & \textcolor{gray!70}{-0.11} & \textcolor{gray!70}{---}                      \\
        34    & \ding{51} & ammonium phosphate monobasic                              & \ce{NH4H2PO4}                                      & 7722-76-1                       & \texttt{216003}                      & 0.98                       & 190                       & \textsuperscript{a}                      & +0.01                      & Speedmix (dry): 20\,min 2500\,rpm                \\
        35    & \ding{51} & sodium carbonate                                          & \ce{Na2CO3}                                        & 497-19-8                        & \texttt{222321}                      & 0.995                      & 856                       & \textsuperscript{h}                      & +0.12                      & Mortar \& pestle 5\,min                       \\
        36    & --        & \textcolor{gray!70}{sodium phosphate monobasic dihydrate} & \textcolor{gray!70}{\ce{NaH2PO4}$\cdot$\ce{2H2O}}  & \textcolor{gray!70}{13472-35-0} & \textcolor{gray!70}{\texttt{71500}}  & \textcolor{gray!70}{0.99}  & \textcolor{gray!70}{225}   & \textcolor{gray!70}{\textsuperscript{c}} & \textcolor{gray!70}{+2.7}  & \textcolor{gray!70}{Mortar \& pestle 10\,min} \\
        37    & --        & \textcolor{gray!70}{sodium nitrate}                       & \textcolor{gray!70}{\ce{NaNO3}}                    & \textcolor{gray!70}{7631-99-4}  & \textcolor{gray!70}{\texttt{221341}} & \textcolor{gray!70}{0.99}  & \textcolor{gray!70}{307}  & \textcolor{gray!70}{\textsuperscript{a}} & \textcolor{gray!70}{-0.09} & \textcolor{gray!70}{---}                      \\
        38    & \ding{51} & niobium(v) oxide                                          & \ce{Nb2O5}                                         & 1313-96-8                       & \texttt{208515}                      & 0.999                      & 1512                      & \textsuperscript{a}                      & -0.04                      & ---                                           \\
        39    & \ding{51} & nickel(ii) oxide                                          & \ce{NiO}                                           & 1313-99-1                       & \texttt{399523}                      & 0.99                       & 1955                      & \textsuperscript{b}                      & +0.03                      & ---                                           \\
        40    & \ding{51} & lead(ii) oxide                                            & \ce{PbO}                                           & 1317-36-8                       & \texttt{211907}                      & 0.999                      & 886                       & \textsuperscript{a}                      & +0.34                      & ---                                           \\
        41    & \ding{51} & antimony(iii) oxide                                       & \ce{Sb2O3}                                         & 1309-64-4                       & \texttt{230898}                      & 0.99                       & 655                       & \textsuperscript{a}                      & -0.18                      & ---                                           \\
        42    & --        & \textcolor{gray!70}{silicon}                              & \textcolor{gray!70}{\ce{Si}}                       & \textcolor{gray!70}{7440-21-3}  & \textcolor{gray!70}{\texttt{215619}} & \textcolor{gray!70}{0.99}  & \textcolor{gray!70}{1412} & \textcolor{gray!70}{\textsuperscript{a}} & \textcolor{gray!70}{-0.12} & \textcolor{gray!70}{---}                      \\
        43    & \ding{51} & silicon dioxide                                           & \ce{SiO2}                                          & 60676-86-0                      & \texttt{342890}                      & 0.995                      & 1710                      & \textsuperscript{d}                      & +0.18                      & ---                                           \\
        44    & \ding{51} & tin(iv) oxide                                             & \ce{SnO2}                                          & 18282-10-5                      & \texttt{244651}                      & 0.999                      & 1630                      & \textsuperscript{a}                      & +0.02                      & ---                                           \\
        45    & \ding{51} & strontium carbonate                                       & \ce{SrCO3}                                         & 1633-05-2                       & \texttt{472018}                      & 0.999                      & 1494                      & \textsuperscript{b}                      & +0.06                      & ---                                           \\
        46    & \ding{51} & titanium(iv) oxide, anatase                               & \ce{TiO2}                                          & 1317-70-0                       & \texttt{232033}                      & 0.998                      & 1843                      & \textsuperscript{a}                      & -0.06                      & ---                                           \\
        47    & \ding{51} & vanadium(iii) oxide                                       & \ce{V2O3}                                          & 1314-34-7                       & \texttt{215988}                      & 0.98                       & 1940                      & \textsuperscript{a}                      & +0.04                      & ---                                           \\
        48    & \ding{51} & vanadium(v) oxide                                         & \ce{V2O5}                                          & 1314-62-1                       & \texttt{223794}                      & 0.98                       & 690                       & \textsuperscript{b}                      & +0.06                      & ---                                           \\
        49    & \ding{51} & tungsten(vi) oxide                                        & \ce{WO3}                                           & 1314-35-8                       & \texttt{95410}                       & 0.999                      & 1473                      & \textsuperscript{b}                      & -0.14                      & High energy ball mill: 10\,min                 \\
        50    & \ding{51} & yttrium(iii) oxide                                        & \ce{Y2O3}                                          & 1314-36-9                       & \texttt{205168}                      & 0.9999                     & 2440                      & \textsuperscript{a}                      & +0.05                      & ---                                           \\
        51    & \ding{51} & zinc oxide                                                & \ce{ZnO}                                           & 1314-13-2                       & \texttt{96479}                       & 0.99                       & 1975                      & \textsuperscript{a}                      & -0.16                      & ---                                           \\
        52    & \ding{51} & zirconium(iv) oxide                                       & \ce{ZrO2}                                          & 1314-23-4                       & \texttt{230693}                      & 0.99                       & 2715                      & \textsuperscript{b}                      & +0.14                      & ---                                           \\
        \bottomrule
    \end{tabularx}
\end{sidewaystable}

\subsection{Experimental Synthesis}

All solid-state reactions are performed using the A-Lab, an autonomous solid-state synthesis platform whose infrastructure is described in full by Szymanski \textit{et al.} \cite{Alab}. Briefly, the A-Lab carries out each reaction in three stages that mirror manual synthesis: (\textit{i}) precursor preparation, (\textit{ii}) heating, and (\textit{iii}) sample recovery and characterization. At each stage, a 6-axis robot arm programmed with high-precision waypoints transfers samples between devices orchestrated by AlabOS \cite{Alab_os}.

For each pairwise reaction, the two precursors are combined in stoichiometric amounts chosen to give a 1:1 ratio of the two metal cations, targeting a hypothetical binary product compound; the precise precursor mole ratios used for every reaction are recorded in the \texttt{target\_stoichiometry} field of the ledger.

To prepare samples, powders are dispensed and weighed simultaneously in air by using an automatic dispenser balance (Quantos, Mettler Toledo) within a 0.2\% weight threshold into a plastic vial containing 10 zirconia balls. Weights for each precursor are automatically recorded. Ethanol is added and the sample is mixed at 2000\,rpm for 10 minutes to create a slurry. The slurry is pipetted from the vial and dispensed into a crucible. After drying at 80$^\circ$C, the sample is left mixed and densified at the bottom of a crucible. The samples are then moved to furnaces and heated according to the programmed profile. The default heating profile is 2\,$^\circ$C/min to 300\,$^\circ$C, followed by 5\,$^\circ$C/min up to the maximum dwell temperature, followed by a 1-hour hold, and then uncontrolled cooling in the furnace to room temperature. A thermocouple (Barnstead Thermocouple Assembly type K) recorded temperature points throughout heating and cooling at approximately minute intervals.

Finally, the samples are recovered autonomously by vertically shaking a crucible covered with a plastic cap with a 10-mm alumina ball inside to dislodge (mill) the sample from the crucible wall and generate powder.

Recovered powder weight is automatically recorded with a Ohaus SPX422 AM.

Finally, the sample is shaken through a steel mesh and pressed onto a disc for automated powder XRD characterization, and the remaining samples are stored inside a uniquely-labeled plastic vial for further characterization.

The reaction temperature is set by Tammann's rule \cite{Tammann_rule_3, Tammann_rule_1}, which prescribes a solid-state synthesis temperature of two-thirds the lowest melting point ($T_m$) among the precursors. Because the A-Lab executes reactions in batched heating runs, the dwell temperature $T_t$ is obtained from Tammann's rule and rounded down to the nearest hundred degrees Celsius:
\begin{equation}
    T_t = 100 \left\lfloor \frac{1}{100} \cdot \frac{2}{3} \min\left( T_{m,1},\, T_{m,2} \right) \right\rfloor.
\end{equation}
The minimum permitted dwell temperature is 200\,$^\circ$C and the maximum is 1,100\,$^\circ$C, bounded by furnace constraints. Melting points used to evaluate the Tammann-rule temperature and their bibliographic sources are provided in Table~\ref{tab:precursor-purchase-info}. Several of the selected precursors decompose or dehydrate before melting (e.g., iron oxalate dihydrate); for these, we used the highest relevant decomposition temperature in place of $T_m$ when calculating the dwell temperature $T_t$. This is an approximation: kinetic heuristics like the Tammann rule implicitly assume a stable compound up to $T_m$ and may not strictly apply to materials that decompose or melt incongruently before they react.

Sample characterization is performed primarily by powder X-ray diffraction (XRD) on an Aeris Minerals diffractometer (Malvern Panalytical) using Cu K$\alpha$ radiation. Scans cover the range 10--100$^\circ$ $2\theta$ over 8\,min, with a step size of $\approx$0.01$^\circ$ and an active length of 5.542$^\circ$.

Scanning electron microscopy (SEM) and energy-dispersive X-ray spectroscopy (EDS) measurements of precursors are taken using a Thermofisher Phenom XL G2 system.

\subsection{Phase Identification and Refinement}

XRD patterns are analyzed by Rietveld refinement using Dara, an in-house automated phase-identification and refinement pipeline built on the BGMN kernel \cite{Dara, BGMN}, and every refinement is subsequently verified by a human expert. Dara ranks candidate phase sets by a quality-of-fit figure of merit, drawing initial candidate structures from the Inorganic Crystal Structure Database (ICSD) and the Crystallography Open Database (COD) \cite{ICSD}. Lattice parameters are allowed to vary by 1\% where symmetry permits. A fourth-order spherical-harmonics preferred-orientation model (\texttt{gewicht=SPHAR4}) is used throughout refinement. The crystallite-size parameters $k_1$ and $B_1$ are constrained isotropically to $[0,\,0.1]$ with an initial value of zero, and the microstrain parameter $k_2$ is constrained to $[0,\,0.1]$.

\subsection{Computational Thermodynamics Calculations}

Reaction energies are computed by querying the Materials Project \cite{MaterialsProject} for DFT total energies through \texttt{pymatgen} \cite{pymatgen}. These energies are then used to construct phase diagrams \cite{MaterialsProject_PhaseDiagram}, reaction interfaces \cite{MaterialsProject_ReactionInterface1, MaterialsProject_ReactionInterface2}, and reaction networks \cite{McDermott_reactionNetwork_model}, evaluated both at 0\,K and at the Tammann-rule temperature using a physical-descriptor fit for the Gibbs energy \cite{Gibbs_Bartel}.

\section{Data Records}

\begin{figure}[ht!]
    \centering
    \includegraphics[width=1.0\linewidth]{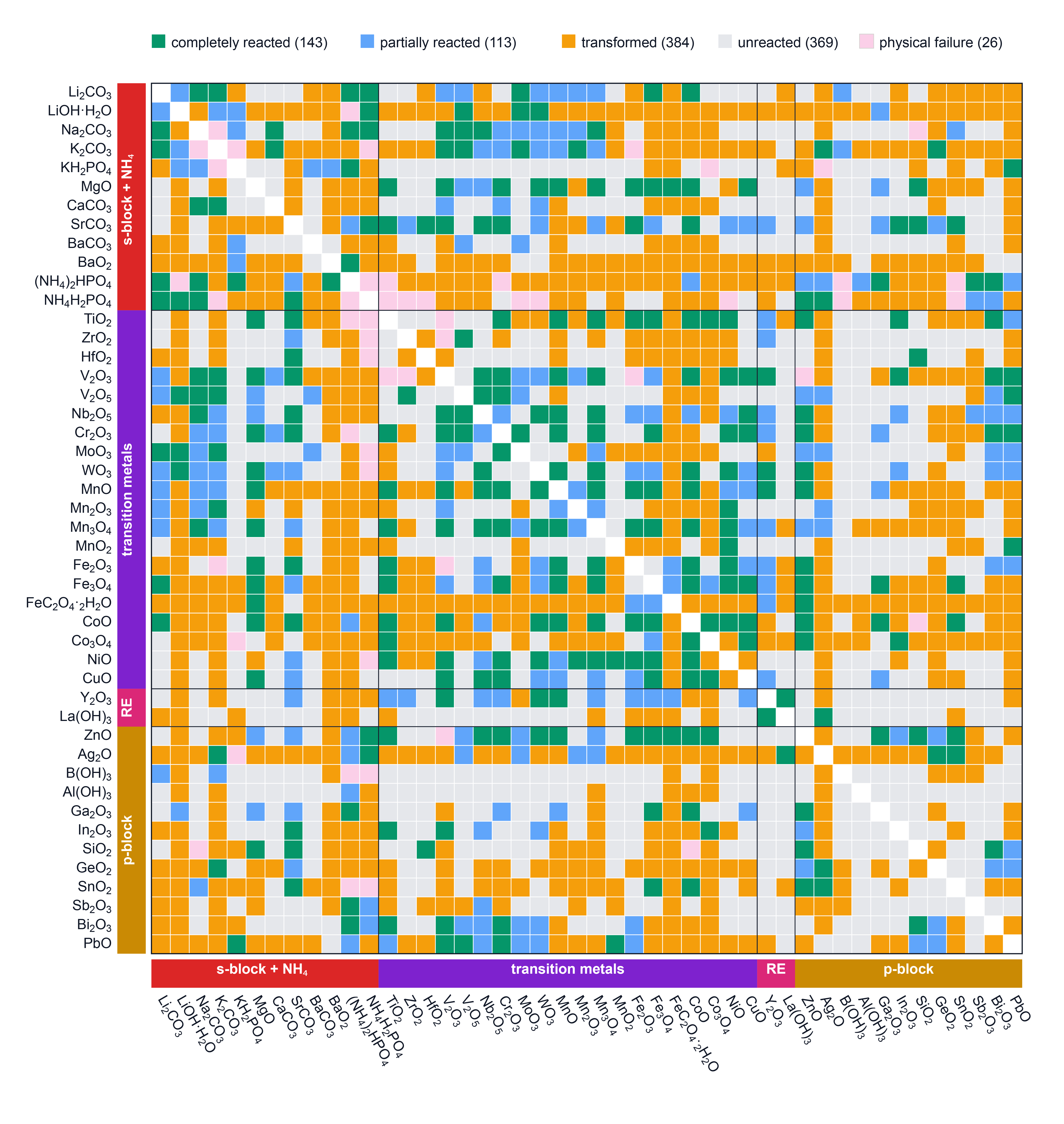}
    \caption{\textbf{Pairwise reaction coverage across the Precursor Genome.} Each cell represents the solid-state reaction between the corresponding row and column precursors, colored by the expert-validated reaction outcome. \textbf{Completely reacted} (emerald) denotes that the target phase was formed with no detectable residual precursors. \textbf{Partially reacted} (blue) denotes that the target phase was detected alongside remaining precursor phases. \textbf{Transformed} (amber) denotes that a precursor decomposed or reacted to form a variant precursor phase (e.g., a carbonate losing \ce{CO2} to its parent oxide, or \ce{Ag2O} oxidizing), but the target phase did not form. \textbf{Unreacted} (gray) denotes that only the original precursor phases were observed.
    \textbf{Physical failure} (light pink) denotes samples unable to be handled automatically post-heating due physical incompatibility (e.g. liquid or clay-like sample).
    %
    Colored sidebars group the 46 precursors into four chemical families---s-block and ammonium salts (red), transition metals (purple), group 3 and rare earth (pink), and p-block (gold)---and thin rules inside the matrix mark the boundaries between these families.}
    \label{fig:coverage-matrix}
\end{figure}

\begin{figure}[ht!]
    \centering
    \includegraphics[width=0.6\linewidth]{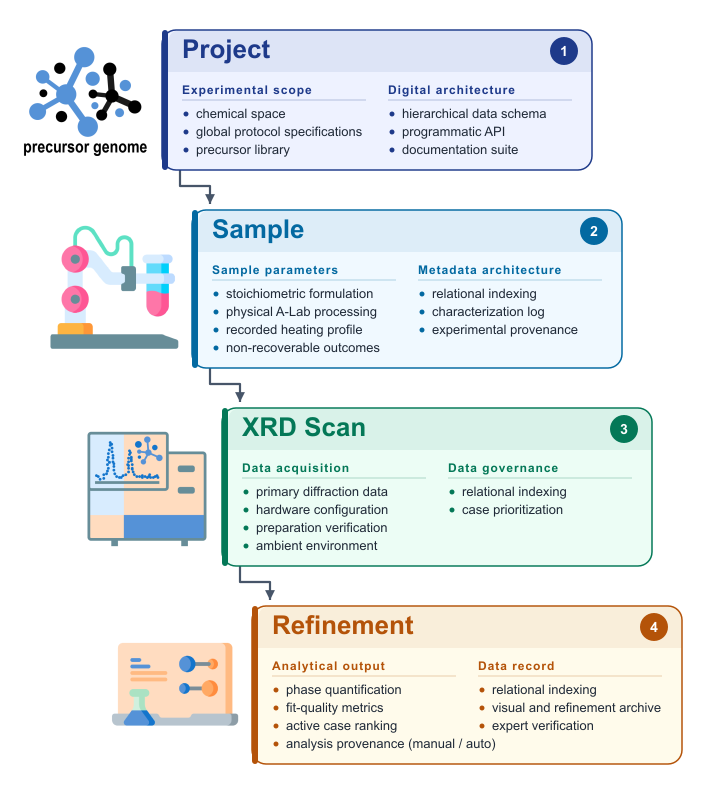}
    \caption{\textbf{The Precursor Genome data hierarchy.} The Pydantic schema is organized as four color-coded tiers---\textbf{Project}, \textbf{Sample}, \textbf{XRD Scan}, and \textbf{Refinement}---each split into two paired columns that separate experimental or analytical content from the associated metadata and digital architecture. \textbf{Project} defines the chemical-space boundaries, precursor library, and global protocol specifications, alongside the digital artifacts (hierarchical schema, programmatic API, documentation suite) that make those specifications machine-readable. \textbf{Sample} houses the stoichiometric formulation, physical A-Lab processing parameters, the measured heating profile, and non-recoverable outcomes, alongside the relational indexing, characterization log, and experimental provenance used to join sample-level metadata to downstream records. XRD \textbf{Scan} archives the primary diffraction data together with the hardware configuration, preparation-verification audit, and ambient environment under which each measurement was acquired, along with the relational indexing and case prioritization that govern its use. \textbf{Refinement} stores the analytical output for each scan---phase quantification, fit-quality metrics, the active case ranking, and the manual/automatic provenance of every candidate---together with the visual and refinement archive and the expert verification that anchors the final result.}
    \label{fig:database-map}
\end{figure}

\autoref{fig:coverage-matrix} provides a visual summary of the 1,035 reactions contained in the dataset, organized by precursor pair and colored according to the expert-validated reaction outcome. The underlying experimental and characterization output is compiled into a hierarchical repository implemented with Pydantic; \autoref{fig:database-map} shows a schematic of the data structure and contents. The top-level \texttt{Project} class encompasses the whole project and stores the objects and metadata that define its scope, including the precursor library, the full chemical space, and the global protocol (workflow) specifications. 

The dataset is distributed as a custom Pydantic data structure, the organization of which is summarized in Figs.~\ref{fig:schema-experimental} and~\ref{fig:schema-analysis}; it is freely available on GitHub at \href{https://github.com/lauren-walters/precursor-genome.git}{https://github.com/lauren-walters/precursor-genome.git} and archived in full on Zenodo at \href{https://doi.org/10.5281/zenodo.21285546}{https://doi.org/10.5281/zenodo.21285546}. The dataset contains the following:

\begin{figure}[!htbp]
    \centering
    \includegraphics[width=0.6\linewidth]{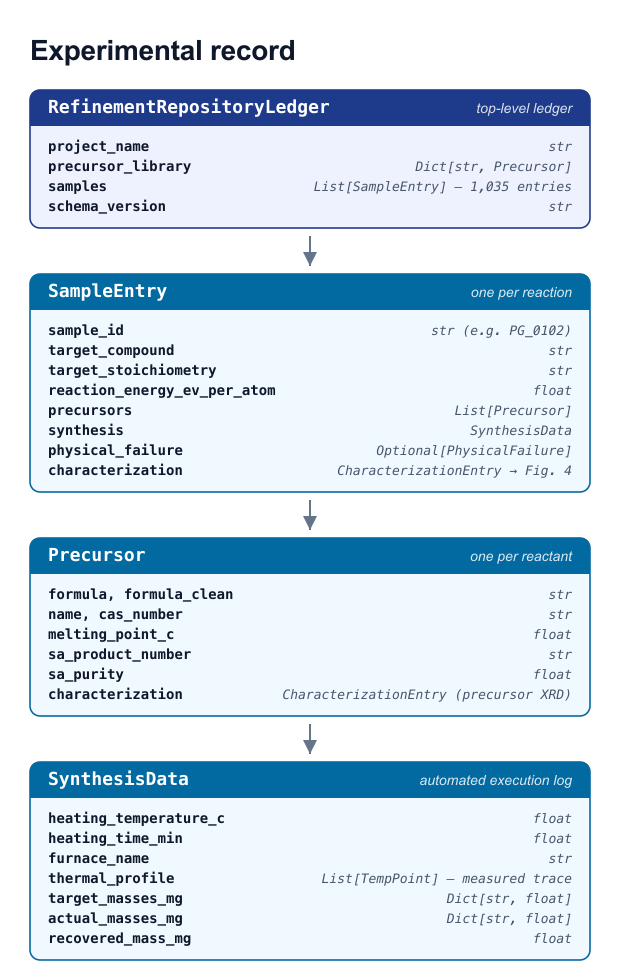}
    \caption{\textbf{Experimental-record half of the Pydantic schema.} The top-level \texttt{RefinementRepositoryLedger} holds 1,035 \texttt{SampleEntry} objects, each of which aggregates its reactants (\texttt{Precursor}) and the automated execution log (\texttt{SynthesisData}, including measured thermal profile and precursor/recovered masses). The analysis half of the schema is continued in \autoref{fig:schema-analysis}.}
    \label{fig:schema-experimental}
\end{figure}

\begin{figure}[!htbp]
    \centering
    \includegraphics[width=0.6\linewidth]{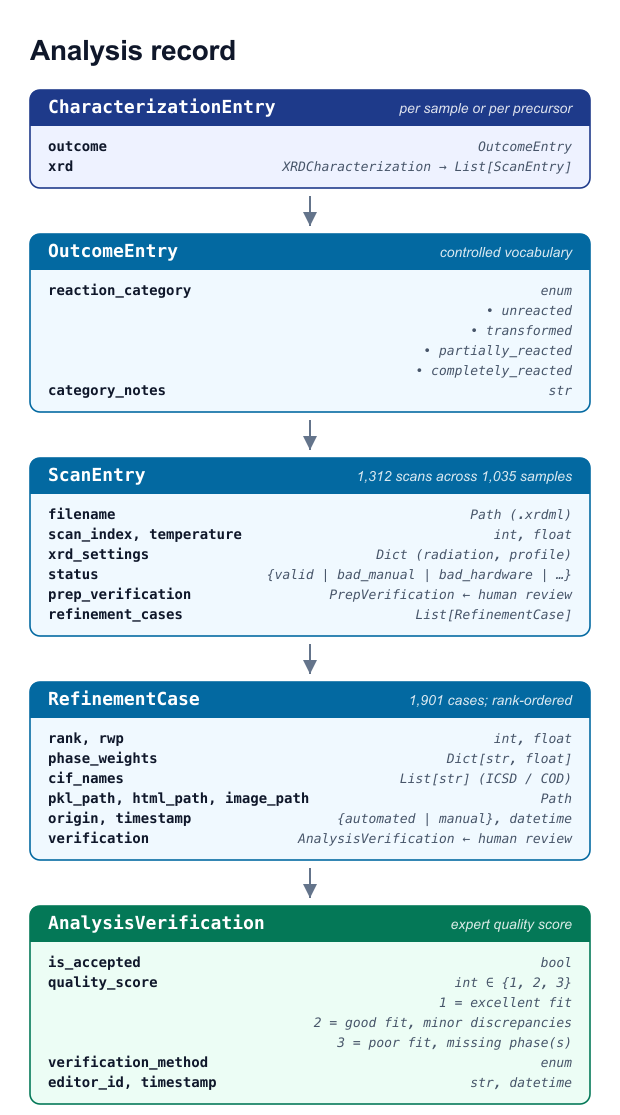}
    \caption{\textbf{Analysis-record half of the Pydantic schema.} Continuing from \autoref{fig:schema-experimental}, each sample's \texttt{CharacterizationEntry} flows through the controlled reaction-outcome vocabulary (\texttt{OutcomeEntry}), raw diffraction scans (\texttt{ScanEntry}, 1,351 total), and rank-ordered Rietveld candidates (\texttt{RefinementCase}, 1,950 total) to the final \texttt{AnalysisVerification} record that captures the 1--3 expert quality score. Cards shaded green indicate fields populated by human review; all other cards are populated automatically by the A-Lab pipeline.}
    \label{fig:schema-analysis}
\end{figure}

\subsection{Precursor Genome Ledger}
The primary data record is \texttt{ledger\_precursor\_genome.json}, a structured JSON file containing 1,035 solid-state synthesis experiments performed by the A-Lab self-driving laboratory. The ledger is organized as an \texttt{ExperimentRepositoryLedger} object containing a list of \texttt{SampleEntry} records. The schema is defined in \texttt{schemas\_precursor\_genome\_v5.py} using the Pydantic data validation library. The top-level ledger fields are:

\begin{itemize}
    \item \texttt{repository\_name} \textit{str}. Name of the repository.
    \item \texttt{sample\_dir} \textit{Path}. Root directory under which all sample files are stored.
    \item \texttt{xrd\_settings} \textit{Dict}. Default instrument configuration applied to all scans, including radiation source (Cu K$\alpha$) and instrument profile (Aeris-fds-Pixcel1d-Medipix3).
    \item \texttt{workflow\_spec} \textit{Optional[WorkflowSpec]}. Project-level protocol specification describing the standard experimental workflow; see the \texttt{WorkflowSpec} entry below.
    \item \texttt{samples} \textit{List[SampleEntry]}. The full list of 1,035 reaction records; see the \texttt{SampleEntry} entry below.
\end{itemize}

\subsubsection*{\texttt{SampleEntry}}
Each \texttt{SampleEntry} is a single row of the ledger and corresponds to one solid-state reaction campaign. Its fields are:
\begin{itemize}
    \item \textbf{\texttt{sample\_id}} \emph{str.} Unique identifier for the experiment (e.g., \texttt{PG\_0102}).
    \item \textbf{\texttt{description}} \emph{Optional[str].} Free-text description of the sample.
    \item \textbf{\texttt{precursors}} \emph{Optional[List[Precursor]].} The reactants used, each recording chemical formula, common name, CAS number, Sigma-Aldrich product number, purity, and melting point; see the \texttt{Precursor} entry below.
    \item \textbf{\texttt{target\_compound}} \emph{Optional[str].} Intended product formula.
    \item \textbf{\texttt{target\_stoichiometry}} \emph{Optional[str].} Precursor stoichiometric ratio (e.g., \ce{Ag2O}--\ce{2Al(OH)3}).
    \item \textbf{\texttt{reaction\_energy\_ev\_per\_atom}} \emph{Optional[float].} DFT-computed reaction energy in eV/atom for the target reaction, evaluated at 0\,K and at the Tammann-rule temperature.
    \item \textbf{\texttt{active\_scan\_index}} \emph{Optional[int].} Index into the list of XRD scans identifying the primary scan selected for this sample.
    \item \textbf{\texttt{physical\_failure}} \emph{Optional[PhysicalFailure].} Record of any mechanical failure of the sample during handling, including a free-text observation and a provenance source.
    \item \textbf{\texttt{synthesis}} \emph{Optional[SynthesisData].} Experimental synthesis conditions, including heating temperature ($^\circ$C), heating time (min), furnace name, full temperature profile, target and actual precursor masses (mg), recovered mass (mg), mass used for XRD (mg), and a flag indicating whether the target mass was met.
    \item \textbf{\texttt{outcome}} \emph{Optional[OutcomeEntry].} Sample-level reaction outcome assigned during expert categorization; see the \texttt{OutcomeEntry} entry below.
    \item \textbf{\texttt{characterization}} \emph{CharacterizationEntry.} All characterization performed on the sample; see the \texttt{CharacterizationEntry} entry below.
\end{itemize}

\subsubsection*{\texttt{Precursor}}
\noindent Each \texttt{Precursor} record captures a single starting material:
\begin{itemize}
    \item \textbf{\texttt{formula}} \emph{str, required.} Chemical formula of the precursor material.
    \item \textbf{\texttt{formula\_clean}} \emph{Optional[str].} Cleaned and standardized version of the formula.
    \item \textbf{\texttt{name}} \emph{Optional[str].} Common or commercial name of the material.
    \item \textbf{\texttt{cas\_number}} \emph{Optional[str].} Chemical Abstracts Service (CAS) registry number.
    \item \textbf{\texttt{melting\_point\_c}} \emph{Optional[float].} Melting point in $^\circ$C.
    \item \textbf{\texttt{sa\_product\_number}} \emph{Optional[str].} Sigma-Aldrich product number.
    \item \textbf{\texttt{sa\_purity}} \emph{Optional[float].} Purity as reported by Sigma-Aldrich.
    \item \textbf{\texttt{characterization}} \emph{Optional[CharacterizationEntry].} Any characterization performed directly on the precursor material itself, including XRD, SEM, and EDS measurements, structured identically to the sample-level characterization so that precursor phase purity is verified and stored within the same data structure.
    \item \textbf{\texttt{preparation\_notes}} \emph{Optional[str].} Free-text notes on any preprocessing applied to the precursor prior to synthesis (e.g., milling).
    \item \textbf{\texttt{metadata}} \emph{Dict.} Flexible key--value store for additional precursor-level information.
\end{itemize}

\subsubsection*{\texttt{CharacterizationEntry}}
\noindent Each \texttt{CharacterizationEntry} aggregates the characterization performed on a sample (or, when nested under a \texttt{Precursor}, on a starting material):
\begin{itemize}
    \item \textbf{\texttt{xrd}} \emph{Optional[XRDCharacterization].} XRD characterization of the sample, including an ordered list of \texttt{ScanEntry} records and an XRD-level \texttt{OutcomeEntry}; see the \texttt{XRDCharacterization} entry below.
    \item \textbf{\texttt{sem}} \emph{Optional[SemCharacterization].} SEM characterization of the sample, including image paths, preparation notes, and an optional outcome record; see the \texttt{SemCharacterization} entry below.
    \item \textbf{\texttt{eds}} \emph{Optional[EdsCharacterization].} EDS characterization of the sample, including image paths, quantification tables, raw spectra, preparation notes, and an optional outcome record; see the \texttt{EdsCharacterization} entry below.
\end{itemize}

\subsubsection*{\texttt{OutcomeEntry}}
\noindent Each \texttt{OutcomeEntry} records the interpreted result of a characterization measurement using a controlled reaction-outcome vocabulary:
\begin{itemize}
    \item \textbf{\texttt{reaction\_category}} \emph{Optional[str].} Controlled-vocabulary reaction category---one of \texttt{unreacted} (no reaction observed; only precursor phases remain), \texttt{transformed} (precursor phases partially consumed; new non-target phases present), \texttt{partially\_reacted} (target phase formed alongside remaining precursors), or \texttt{completely\_reacted} (target phase formed with no detectable residual precursors).
    \item \textbf{\texttt{category\_notes}} \emph{Optional[str].} Free-text notes reporting the specific phases identified, their weight fractions, and any relevant observations.
    \item \textbf{\texttt{phases}} \emph{List[Dict].} Phase composition snapshot derived from the best refinement case, recorded as a list of name--weight-percent pairs (e.g., \texttt{[\{"name": "Ag2O\_224\_...", "weight\_percent": 45.2\}]}).
    \item \textbf{\texttt{phases\_unavailable\_reason}} \emph{Optional[str].} Free-text explanation for why phase data could not be populated, if applicable.
\end{itemize}

\subsubsection*{\texttt{XRDCharacterization}}
\noindent Each \texttt{XRDCharacterization} record aggregates the X-ray diffraction measurements performed on a sample or precursor material:
\begin{itemize}
    \item \textbf{\texttt{scans}} \emph{List[ScanEntry].} Ordered list of XRD measurements performed on the sample, each corresponding to a single diffraction pattern; see the \texttt{ScanEntry} entry below.
    \item \textbf{\texttt{outcome}} \emph{Optional[OutcomeEntry].} XRD-level characterization outcome, recording the phase composition snapshot derived from the best refinement case. Note that \texttt{reaction\_category} and \texttt{category\_notes} are set at the sample level only and should remain null here; see the \texttt{OutcomeEntry} entry above.
\end{itemize}

\subsubsection*{\texttt{ScanEntry}}
\noindent Each \texttt{ScanEntry} represents a single XRD measurement:
\begin{itemize}
    \item \textbf{\texttt{filename}} \emph{str.} Path to the raw \texttt{.xrdml} diffraction file.
    \item \textbf{\texttt{scan\_index}} \emph{int.} Integer index of the scan within the sample.
    \item \textbf{\texttt{temperature}} \emph{Optional[float].} Measurement temperature in $^\circ$C.
    \item \textbf{\texttt{xrd\_settings}} \emph{Dict.} Instrument configuration, including radiation source (Cu K$\alpha$) and instrument profile (Aeris-fds-Pixcel1d-Medipix3).
    \item \textbf{\texttt{status}} \emph{enum.} Scan quality status---one of \texttt{valid}, \texttt{bad\_manual}, \texttt{bad\_hardware}, \texttt{failed\_auto}, or \texttt{superseded}.
    \item \textbf{\texttt{is\_active}} \emph{bool.} Whether this scan is the primary scan for the sample.
    \item \textbf{\texttt{prep\_verification}} \emph{Optional[PrepVerification].} Record documenting whether the physical sample preparation was acceptable, including a boolean acceptance flag and the verification method (\texttt{manual}, \texttt{automated}, or \texttt{pending}).
    \item \textbf{\texttt{automated\_attempted}} \emph{bool.} Whether automated refinement was attempted.
    \item \textbf{\texttt{active\_case\_index}} \emph{Optional[int].} Index into \texttt{refinement\_cases} for the currently accepted case.
    \item \textbf{\texttt{refinement\_cases}} \emph{List[RefinementCase].} Rank-ordered list of candidate refinement results; see the \texttt{RefinementCase} entry below.
\end{itemize}

\subsubsection*{\texttt{RefinementCase}}
\noindent Each \texttt{RefinementCase} represents a single candidate Rietveld refinement produced by the Dara framework:
\begin{itemize}
    \item \textbf{\texttt{rank}} \emph{int.} Integer rank of this refinement candidate (1 = best); manual refinements are assigned rank $-1$ and are always preferred over automated candidates.
    \item \textbf{\texttt{rwp}} \emph{float.} Weighted profile $R$-factor (goodness of fit).
    \item \textbf{\texttt{phase\_weights}} \emph{Dict[str, float].} Mapping of CIF phase names to their refined weight fractions.
    \item \textbf{\texttt{cif\_names}} \emph{List[str].} CIF structure names included in the refinement.
    \item \textbf{\texttt{pkl\_path}} \emph{str.} Path to the serialized refinement object (\texttt{.pkl}).
    \item \textbf{\texttt{html\_path}} \emph{str.} Path to the interactive refinement plot (\texttt{.html}).
    \item \textbf{\texttt{image\_path}} \emph{str.} Path to the static refinement plot (\texttt{.png}).
    \item \textbf{\texttt{origin}} \emph{str.} Provenance of the refinement---\texttt{automated} or \texttt{manual}.
    \item \textbf{\texttt{verification}} \emph{Optional[AnalysisVerification].} Human review record for this refinement, comprising the following fields:
    \begin{itemize}
        \item \textbf{\texttt{is\_accepted}} \emph{bool.} Whether the refinement has been accepted by a human expert.
        \item \textbf{\texttt{method}} \emph{enum.} Verification method---one of \texttt{manual}, \texttt{automated}, or \texttt{pending}.
        \item \textbf{\texttt{editor\_id}} \emph{str.} Identifier of the reviewer who performed the verification.
        \item \textbf{\texttt{quality\_score\_history}} \emph{List[int].} Full history of quality scores assigned by human reviewers (1 = excellent fit; 2 = good fit with minor discrepancies; 3 = poor fit with missing phases).
        \item \textbf{\texttt{human\_quality\_score}} \emph{Optional[int], computed.} Modal quality score derived from \texttt{quality\_score\_history}; \texttt{None} if no scores have been recorded. When the first two reviewer evaluations disagreed, a third expert arbitrated and the modal score of the two matching evaluations was assigned.
    \end{itemize}
\end{itemize}

\subsubsection*{\texttt{EdsCharacterization}}
\noindent Each \texttt{EdsCharacterization} record aggregates the energy-dispersive X-ray spectroscopy measurements performed on a sample or precursor material:
\begin{itemize}
    \item \textbf{\texttt{images}} \emph{List[str].} Paths to EDS map images.
    \item \textbf{\texttt{csv\_files}} \emph{List[str].} Paths to quantification tables in CSV format.
    \item \textbf{\texttt{emsa\_files}} \emph{List[str].} Paths to raw EDS spectra in EMSA format.
    \item \textbf{\texttt{outcome}} \emph{Optional[OutcomeEntry].} EDS-level characterization outcome; see the \texttt{OutcomeEntry} entry above.
    \item \textbf{\texttt{preparation\_notes}} \emph{Optional[str].} Free-text notes on sample preparation prior to EDS measurement.
    \item \textbf{\texttt{metadata}} \emph{Dict.} Flexible key--value store for additional EDS-level information.
\end{itemize}

\subsubsection*{\texttt{SemCharacterization}}
\noindent Each \texttt{SemCharacterization} record aggregates the scanning electron microscopy measurements performed on a sample or precursor material:
\begin{itemize}
    \item \textbf{\texttt{images}} \emph{List[str].} Paths to SEM images.
    \item \textbf{\texttt{outcome}} \emph{Optional[OutcomeEntry].} SEM-level characterization outcome; see the \texttt{OutcomeEntry} entry above.
    \item \textbf{\texttt{preparation\_notes}} \emph{Optional[str].} Free-text notes on sample preparation prior to SEM measurement.
    \item \textbf{\texttt{metadata}} \emph{Dict.} Flexible key--value store for additional SEM-level information.
\end{itemize}

\subsection{Raw X-ray Diffraction Data}
Raw diffraction patterns are provided in XRDML (.xrdml) format as generated by a Malvern Panalytical Aeris diffractometer using Cu K$\alpha$ radiation and a Pixcel1D-Medipix3 detector. The dataset contains 1,351 scans across 1,035 samples. All scans have a status of valid in the ledger.

\subsection{Rietveld Refinement Results}
Automated Rietveld refinement results are provided for 1,950 refinement cases across 1,351 scans. Each case is extracted from a Dara Refinement Object, and includes refined phase weight fractions, CIF phase identifications, and weighted profile R-factor (Rwp) values. Refinement objects are stored as serialized Python objects in PKL (.pkl) format. Interactive refinement plots generated using Plotly are provided in HTML (.html) format, and static images are provided in PNG (.png) format.

\section{Technical Validation}

Technical validation of the synthesis metadata was maintained through systematic data acquisition and instrument calibration within the A-Lab \cite{Alab, Alab_os}. Continuous parameters, such as the thermal profile, were logged at regular intervals to ensure fidelity of the recorded trajectory, while discrete measurements, such as precursor mass, were recorded on analytical balances that were periodically calibrated and manually verified. To minimize human error, all experimental metadata were captured automatically by the autonomous system. To ensure operational consistency, every automated sample-handling device underwent soak testing over a minimum of 25 iterations prior to data collection, and automated error handling was employed wherever possible to minimize human intervention. To prevent cross-contamination, all chemical reagents were dedicated exclusively to this project.

Reaction results were validated through a multi-stage process combining automated refinement with extensive human expert oversight. All diffraction patterns were initially processed via the Dara framework \cite{Dara}, which first identified potential phases from a comprehensive database using peak matching. Structural information (CIF files) was sourced from the ICSD \cite{ICSD} and COD \cite{COD_1, COD_2, COD_3, COD_4, COD_5, COD_6, COD_7, COD_8, COD_9}. Subsequently, a tree-search algorithm enumerated and combined candidate phases, selecting the optimal configuration based on numerical goodness of fit ($R_{wp}$) and qualitative metrics, such as missing or extra calculated peaks.
Following these automated steps, researchers performed a verification pass to confirm or reject the proposed refinements. During this stage, scans with an insufficient signal-to-noise ratio were flagged for re-acquisition. For any cases where the automated selection was deemed incorrect by a human expert, a manual refinement was performed to ensure phase identification accuracy.

Final quality scores were assigned by human experts to quantify the reliability of each refinement. Because our two reference databases cannot cover all materials, and because real phases often deviate from their documented forms, a small fraction of candidate phases could not be reliably identified during refinement. To account for this, we established a three-tier quality metric: (1) an excellent fit with high trustworthiness; (2) a good fit with minor discrepancies, such as one or two small missing peaks or intensity mismatches; and (3) a poor fit characterized by at least one large missing phase and/or significant unfit intensity.
Each scan was evaluated independently by two human experts. When the first two evaluations disagreed, a third expert arbitrated the final score; the final quality designation was then assigned from the two matching evaluations, ensuring a consistent assessment of the data.

\section{Usage Notes}

The Precursor Genome is distributed as a single top-level JSON file (\texttt{ledger\_precursor\_genome.json}) accompanied by the raw and derived characterization artifacts it references by path. The ledger validates against the Pydantic schema shipped in \texttt{schemas\_precursor\_genome.py}, which also serves as executable documentation of every field introduced in Figs.~\ref{fig:schema-experimental} and~\ref{fig:schema-analysis}.

\noindent \textbf{Access.} The complete dataset, including the JSON ledger, raw \texttt{.xrdml} files, serialized refinement objects, and static and interactive refinement plots, is archived on Zenodo at \href{https://doi.org/10.5281/zenodo.21285546}{https://doi.org/10.5281/zenodo.21285546} and is mirrored, together with all loader code, on GitHub at \href{https://github.com/lauren-walters/precursor-genome.git}{https://github.com/lauren-walters/precursor-genome.git}. Both repositories are released under the Creative Commons Attribution 4.0 International (CC BY 4.0) license.

\noindent \textbf{Software requirements.} Loading and querying the ledger requires Python 3.10 or later, \texttt{pydantic} (v2), and \texttt{numpy}. Parsing and plotting raw diffraction patterns additionally requires \texttt{xylib} (for \texttt{.xrdml} ingestion) and either \texttt{pymatgen} or \texttt{GSAS-II} for downstream crystallographic analysis. Interactive refinement plots are served as self-contained HTML files generated with \texttt{plotly}.

\noindent \textbf{Tutorial notebooks.} Three Jupyter notebooks accompany the GitHub repository and provide a practical starting point for working with the ledger. \texttt{01\_load\_and\_explore.ipynb} loads \texttt{ledger\_precursor\_genome.json} through the Pydantic schema, flattens the 1,035 \texttt{SampleEntry} records into a single pandas DataFrame, and reports summary statistics (reaction-category counts, scan and refinement totals) alongside single-sample lookup by \texttt{sample\_id}. \texttt{02\_filter\_and\_query.ipynb} demonstrates filtering by reaction outcome and by the elemental composition of precursors, extracts per-phase weight fractions from each sample's \texttt{best\_refinement} case, and exports filtered subsets to CSV. \texttt{03\_plot\_xrd\_patterns.ipynb} parses raw \texttt{.xrdml} files directly---the bundled patterns use Panalytical schema v2.2/v2.3, not yet supported by the \texttt{xrdtools} library---plots single and overlaid diffraction patterns, and loads a serialized Rietveld refinement object to reproduce its observed/calculated/background fit alongside the corresponding rendered image. Raw \texttt{.xrdml} scans, refinement \texttt{.pkl} objects, and rendered fit images are archived on Zenodo rather than bundled in the GitHub repository; the first two notebooks operate on the ledger alone and are unaffected by this split, while the third detects local data availability and prints an explanatory message in place of any cell that requires files not present locally.

\noindent \textbf{Reaction-outcome vocabulary.} Every scan is annotated with one of four controlled outcomes, ordered by increasing degree of conversion: \texttt{unreacted} (no reaction observed; only precursor phases remain), \texttt{transformed} (precursors partially consumed but no target phase formed), \texttt{partially\_reacted} (target phase formed alongside residual precursor phases), and \texttt{completely\_reacted} (target phase formed with no detectable residual precursors). Downstream users building binary classifiers should decide explicitly whether to treat \texttt{partially\_reacted} as a positive or negative label; we recommend reporting both splits.

\noindent \textbf{Working with quality scores.} Each \texttt{RefinementCase} carries a three-tier expert quality score (1 = excellent, 2 = good with minor discrepancies, 3 = poor). Downstream analyses should either (i) restrict training to quality-1 refinements, (ii) weight samples by inverse quality score, or (iii) use the score as an auxiliary regression target. We strongly discourage silently mixing quality-3 records into quantitative fits without acknowledgment.

\noindent \textbf{Active vs candidate refinements.} Every scan retains the full list of ranked \texttt{RefinementCase} candidates considered by the Dara framework, not only the accepted one. Users who want a single authoritative phase assignment per scan should index into \texttt{refinement\_cases[active\_case\_index]}; users building models that learn to rank candidate phase sets can use the full list together with the \texttt{rank} and \texttt{rwp} fields.

\noindent \textbf{Filtering for physical validity.} Samples that experienced mechanical handling failures during processing carry a non-null \texttt{physical\_failure} field, and scans that failed automated or manual QC carry a \texttt{status} other than \texttt{valid}. Both filters should be applied before using records as ground-truth labels for machine learning.

\noindent \textbf{Citation.} Users of the dataset are asked to cite this article and the Zenodo archive DOI, and, where the analysis depends on automated phase identification, to additionally cite the Dara framework \cite{Dara} and the BGMN refinement kernel \cite{BGMN}.

\section{Data Availability}

The data for all synthesis conditions, raw characterization files, and (semi-)automated characterization is available in full on Zenodo at \href{https://doi.org/10.5281/zenodo.21285546}{https://doi.org/10.5281/zenodo.21285546}.

\section{Code Availability}

The code for accessing all data and a summary ledger is included in the GitHub repository \href{https://github.com/lauren-walters/precursor-genome.git}{https://github.com/lauren-walters/precursor-genome.git} with full data archived on Zenodo at \href{https://doi.org/10.5281/zenodo.21285546}{https://doi.org/10.5281/zenodo.21285546}.

\section{Author Contributions}
Author contributions to dataset include conceptualization (all); synthesis planning (M.J.M, L.N.W., B.R.), synthesis execution (M.J.M, L.N.W., B.R., Y.F.), characterization and analysis (L.N.W., M.J.M, Y.F.), data preparation (L.N.W., B.R., Y.F.), data structure creation (L.N.W, B.R.), manuscript preparation (L.N.W., M.J.M), manuscript revision and publication (all), supervision (G.C.), and funding and resource acquisition (G.C.).

\section{Competing Interests}
The authors declare that they have no competing interests.

\section{Acknowledgments}
This work was funded and intellectually led by the D2S2 program within the U.S. Department of Energy, Office of Basic Energy Sciences, Materials Sciences and Engineering Division under Contract No. DE-AC02-05-CH11231.
L.N.W. acknowledges funding support from the BIDMaP Postdoctoral Fellowship.
B.R. acknowledges support from the Kavli ENSI Graduate Student Fellowship.

\bibliography{sn-bibliography}

\end{document}